%% file: main.tex
\documentclass{article} 
\usepackage{conference,times}

\input{math_commands.tex}

\usepackage{hyperref}
\usepackage{url}

\usepackage{multirow}
\usepackage{booktabs}
\usepackage{arydshln}
\usepackage{graphicx}
\usepackage{algorithm}
\usepackage{algpseudocode}
\usepackage{amssymb}
\usepackage[table]{xcolor}
\usepackage{graphicx}
\usepackage{subcaption}

\newcommand{\seed}{{\texttt{Seed-Q}}}

\newcommand{\dg}{\textbf{\textsuperscript{$\dagger$}}}
\newcommand{\ddg}{\textbf{\textsuperscript{$\ddagger$}}}

\newcommand{\bstars}{\textbf{\textsuperscript{$\bigstar$}}}

\newcommand{\nop}{---}
\newcommand{\nr}{\textit{NR}}

\definecolor{shade}{RGB}{235,244,242}
\newcommand{\ccbb}{\cellcolor{shade}}

\title{Security-Enhanced Seed-Based Weight \\Quantization for Large Language Models}

\author{Qiuyu Ren, Sudipta Paria, Aritra Dasgupta, Swarup Bhunia\\
Department of Electrical and Computer Engineering (ECE)\\
University of Florida\\
Gainesville, FL 32611, USA \\
\texttt{\{qiuyuren,sudiptaparia,aritradasgupta\}@ufl.edu, swarup@ece.ufl.edu}
}

\iclrfinalcopy
\begin{document}

\maketitle

\begin{abstract}
Large language models (LLMs) incur substantial storage, memory-bandwidth and energy costs, motivating compact weight representations. Existing seed-based compression methods reconstruct weights from compact pseudo-random representations but do not explicitly account for the non-uniform sensitivity of model weights. We introduce \texttt{Seed-Q}, a security-enhanced sensitivity-aware seed-based weight compression framework that uses lightweight Linear Feedback Shift Register (LFSR)-based weight generation with non-uniform bit allocation. Our approach assigns larger representation budgets to sensitive weights while aggressively compressing less sensitive regions. Importantly, this non-uniform allocation requires no side-information: the decoder deterministically reconstructs the bit-allocation schedule, with no rung depending on the decoded weights, eliminating the need to store per-block metadata or use calibration data while preserving the baseline coding rate. Experiments across diverse LLMs show that \texttt{Seed-Q} matches 4-bit perplexity of SeedLM with fewer bits, while at the same 4 bits/weight it reduces both perplexity degradation and zero-shot accuracy loss relative to SeedLM. We also show that \texttt{Seed-Q} simultaneously achieves high security against bit-flip attacks on model parameters, as bit corruption affects multiple reconstructed weights, greatly amplifying its impact and making it easier to detect. We further implement \texttt{Seed-Q} in an ASIC-based accelerator and demonstrate modest hardware overhead compared to prior seed-based approaches.
\end{abstract}

\section{Introduction}

The substantial memory and bandwidth demands of large language models (LLMs) pose a key challenge for efficient inference, highlighting the need for compact and effective weight representations. Weight-only post-training quantization (PTQ) reduces this cost by representing model parameters at low precision, with methods such as OPTQ, AWQ and OmniQuant achieving competitive performance at approximately 3-4 bits per weight~\citep{frantar2023gptq,lin2024awq,shao2024omniquant}. However, these methods often require calibration data or additional optimization and still explicitly store the quantized weights.
Autoregressive decoding repeatedly transfers model weights through the memory hierarchy and is often memory-bound, making off-chip DRAM traffic a major contributor to inference latency and energy consumption~\citep{shafipour2025seedlm}.
For LLM inference, the cost of moving weights through the memory hierarchy can substantially exceed the cost of local computation. The energy cost of a DRAM access (1-2 nJ) is approximately a couple of orders-of-magnitude higher than the cost of an internal cache access or functional operation ($\sim$10 pJ)~\citep{horowitz2014computing}.
This large gap makes model compression particularly valuable for memory-bound inference: fewer bits per weight directly reduce storage and off-chip traffic, with corresponding benefits in memory-access energy and bandwidth demand.
While PTQ addresses this problem using low-precision representations, seed-based approaches further reduce memory accesses by reconstructing weight blocks from compact pseudo-random representations.

Seed-based compression offers an approach that substitutes stored weights with compact generative representations. SeedLM~\citep{shafipour2025seedlm} partitions weights into blocks and reconstructs each block from an LFSR-generated pseudo-random basis and a small set of coefficients, thereby trading inexpensive computation for reduced memory traffic. However, SeedLM applies a uniform representation budget across blocks, implicitly treating compression errors at different model locations as equally important. This ignores the strong non-uniformity in weight sensitivity: similar reconstruction errors can have substantially different effects on model quality depending on where they occur. S-Quant~\citep{wang2026squant} improves seed-based compression by adapting the number of generated bases according to the representational complexity of individual blocks. Nevertheless, its allocation is primarily driven by how well a block can be reconstructed rather than how strongly perturbations to that block affect the model's output. Consequently, neither approach directly incorporates model sensitivity when distributing the available representation budget.

We introduce \seed, a sensitivity-aware seed-based weight compression framework that addresses this limitation while preserving the hardware simplicity of LFSR-based reconstruction. \seed~assigns larger representation budgets to sensitive weight regions and more aggressively compresses less sensitive ones under a constrained average coding rate. Unlike conventional adaptive representations that require an explicit configuration for each block, \seed~makes the allocation decoder-reproducible: the decoder deterministically derives each block's representation from stored gains and compact allocation tables. This eliminates per-block allocation metadata and allows sensitivity-aware variable-rate coding without sacrificing its storage benefit.

\textbf{Goals.}
\seed~improves upon prior seed-based compression along three dimensions.

\par
\noindent
\begin{minipage}[t]{.5\linewidth}
\raggedright
\textit{Storage and energy}: sensitivity-aware allocation reaches 3.77 bits/weight vs. 4.00 bits/weight, reducing both model storage and the amount of weight data transferred from memory, which lowers memory-access energy in bandwidth-bound inference.

\textit{Security and integrity}: seed corruption propagates across multiple reconstructed weights, amplifying bit-flip effects and improving their detectability.

\textit{Hardware efficiency}: \seed{} preserves lightweight LFSR-based reconstruction while achieving modest overhead in the ASIC implementation.

\seed{} reduces the degradation introduced by seed-based weight compression while achieving additional storage reduction. Specifically, \seed{} achieves up to 42\% lower perplexity degradation and 55\% lower accuracy degradation, together with up to 6\% lower bits per weight compared to existing seed-based configurations.
\end{minipage}\hfill
\begin{minipage}[t]{.48\linewidth}
\captionsetup{type=figure,font=small,skip=4pt}
\centering
\raisebox{\dimexpr\ht\strutbox-\height\relax}{%
\includegraphics[width=.99\linewidth]{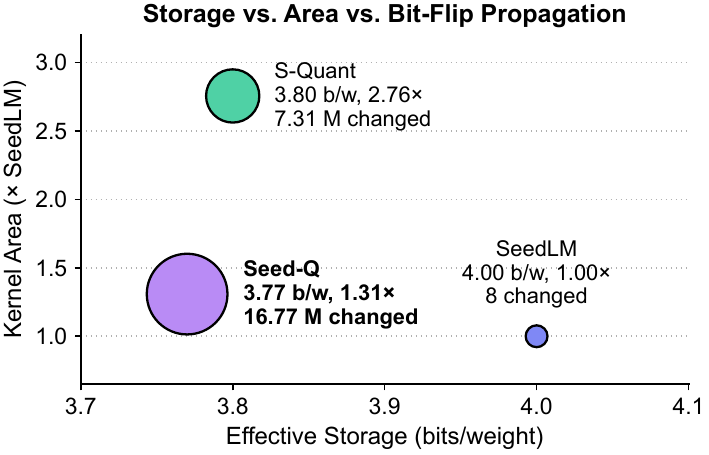}}
\caption{Comparison of \seed~with SeedLM and S-Quant across effective storage, hardware area overhead and bit-flip propagation. The x-axis reports bits per weight, the y-axis shows area normalized to SeedLM and bubble size represents the number of weights affected by a bit flip for Llama-3-8B. \seed~achieves the lowest storage cost with moderate area overhead while exhibiting the largest bit-flip propagation.}
\label{fig:tradeoff}
\end{minipage}
\par

The summary of our contributions is as follows:

\begin{itemize}
    \item We introduce a sensitivity-aware seed-based weight generation scheme that produces quantized weights by modulating a generator function according to sensitivity rather than uniformly or solely according to block reconstruction characteristics.

    \item We present a decoder-reproducible non-uniform allocation that removes the need for allocation metadata, improving the storage-accuracy trade-off of seed-based compression.
    \item We demonstrate its benefits in terms of superior compression quality compared to seed-based weight generation schemes and present the custom hardware (ASIC) implementation overhead compared to existing seed-based techniques.

    \item We present a study on how the proposed scheme can simultaneously provide quantifiable robustness against bit-flip attacks on model weights by making an attack easily detectable.

\end{itemize}

\section{Related Work}

\textbf{Random-basis and seed-based compression.}
Random-basis representations reduce model storage by reconstructing parameters
from compact coefficients and generated bases. PRANC represents network
parameters as linear combinations of pseudo-random basis vectors, while NOLA
applies a related construction to low-rank adaptation matrices~\citep{nooralinejad2022pranc,koohpayegani2023nola}. These approaches employ
relatively large random bases and are not designed for lightweight, on-the-fly
weight generation during LLM inference. SeedLM addresses this limitation by
encoding each weight block using an optimized seed and a small set of
coefficients; a hardware-friendly LFSR regenerates the corresponding basis
during inference, trading inexpensive computation for reduced memory traffic~\citep{shafipour2025seedlm}. S-Quant extends seed-based compression by
adaptively varying the number of generated bases according to the
reconstruction characteristics of individual blocks~\citep{wang2026squant}.
However, SeedLM uses a uniform representation configuration, while S-Quant
primarily adapts capacity according to block reconstructability rather than the
model-level impact of the resulting weight perturbation.

\textbf{Post-training and sensitivity-aware compression.}
PTQ reduces model precision without retraining.
Data-free methods have explored weight equalization, bias correction and
layer-wise compression without access to calibration samples~\citep{nagel2019datafree,horton2020layerwise,nunez2023lcs}. Other approaches
generate synthetic data or use model-generated samples for distillation or
quantization-aware optimization~\citep{lopes2017datafree,gou2021knowledge,liu2023llmqat}.
For LLMs, calibration-based PTQ methods achieve high compression fidelity using
activation or curvature information. OPTQ uses approximate second-order
information for weight quantization, AWQ identifies activation-salient weights,
and OmniQuant optimizes clipping and equivalent transformations~\citep{frantar2023gptq,lin2024awq,shao2024omniquant}. QuIP\# employs incoherence processing and lattice codebooks, while SpQR and AQLM exploit
sparse outlier preservation and additive codebooks, respectively~\citep{tseng2024quipsharp,dettmers2023spqr,egiazarian2024aqlm}.

Several methods further recognize that model parameters have non-uniform
sensitivity to quantization. HAWQ-V2 uses Hessian-based sensitivity for
mixed-precision allocation, while SqueezeLLM combines sensitivity-aware
quantization with the sparse handling of influential weights~\citep{dong2020hawqv2,kim2024squeezellm}. \seed~shares the motivation of
allocating precision non-uniformly but applies it to a different
representation: weights are generated from seeds rather than stored
explicitly as quantized values or codebook indices. Its block-level importance
is derived from checkpoint information rather than token-level activation
calibration and the resulting variable-rate schedule is reproducible by the
decoder without a per-block allocation map.

\textbf{Bit-level integrity of compressed models.}
Prior studies have shown that targeted bit flips in model parameters
can induce substantial accuracy degradation, including through practical memory
fault mechanisms~\citep{rakin2019bitflip,yao2020deephammer}. \seed~considers a
different property of seed-based representations: because one seed contributes
to multiple reconstructed weights, corruption of a seed produces a
multi-weight perturbation.
We measure the fault footprint and its effect on
model quality, which can motivate a future integrity check; we do not evaluate
attack detection.

\section{Methodology}
\label{sec:method}

\seed{} keeps the generation principle of seed-based compression, a
pseudo-random basis expanded from a short LFSR seed, combined with a few
quantized coefficients and changes how many bits each block receives,
choosing for every block one of a small set of seed lengths and coefficient
counts. Blocks can differ without a per-block header because the choice is a
deterministic function of what the decoder already holds: a per-block
importance computed from the checkpoint (\S\ref{sec:importance}), a
rate-distortion allocation over the coding configurations
(\S\ref{sec:alloc}) and a small per-tensor table (\S\ref{sec:decode}). A few
outlier columns that no block format can represent are stored in FP16
(\S\ref{sec:outliers}). Appendix~\ref{app:method} gives the derivations, the
encoder and decoder (Algorithms~\ref{alg:encode} and~\ref{alg:decode}) and the
costs.

\subsection{Seed-based block coding}
\label{sec:codec}

Following SeedLM~\citep{shafipour2025seedlm}, every linear weight matrix
$\mathbf{W}\in\mathbb{R}^{R\times C}$ is split into blocks of $B=8$ weights. A
block $\mathbf{w}$ is represented by an $S$-bit seed $s$ and $k$ quantized
coefficients $\hat{\mathbf{t}}$:
\begin{equation}
  \hat{\mathbf{w}} \;=\; \mathbf{U}(s)\,\hat{\mathbf{t}},
  \qquad \mathbf{U}(s)\in[-1,1]^{B\times k},
  \label{eq:codec}
\end{equation}
where $\mathbf{U}(s)$ is generated by a maximal-length $S$-bit LFSR started at
state $s$ and $\hat{\mathbf{t}}$ quantizes the least-squares coefficients to
$m$-bit mantissas under one shared $e$-bit exponent; the encoder searches all
$2^{S}-1$ seeds for the smallest quantized error. A block costs
\begin{equation}
  \ell(S,k) \;=\; S + m\,k + e \ \text{bits},
  \qquad r(S,k) \;=\; \ell(S,k)/B \ \text{bits per weight},
  \label{eq:rate}
\end{equation}
and SeedLM's 4-bit point is $(S,k)=(16,3)$ with $m=e=4$. We keep SeedLM's
coefficient format throughout, so every difference between the two methods is
in the allocation.

We call $g=(S,k)$ a \emph{rung}; a subset of the grid $S\in\{8,\dots,16\}$,
$k\in\{2,\dots,6\}$ is built for each model, each rung with a fixed rate and,
for a given model, a fixed quality. Blocks could use different rungs; several
seed lengths only need a configurable LFSR polynomial, but a decoder must
know a block's rung before it can find the block's boundaries. Therefore, a per-block choice ordinarily costs $\lceil\log_2 J\rceil$ bits per block with a
fixed-length code over $J$ rungs, $0.25$-$0.375$ bits per weight here. This
is why SeedLM fixes one rung per model. \seed{} removes the per-block cost rather
than the adaptivity.

\subsection{Problem statement}
\label{sec:problem}

Let $g_b$ be the rung of block $b\in\mathcal{B}$. Under the standard local
diagonal-Fisher approximation, perturbing the weights increases the loss by
about $\tfrac12\sum_{r,c}F_{rc}\,\Delta W_{rc}^2$, which suggests scoring a
block by its importance $\mathrm{imp}_b$, the mean of $F$ over the block,
times the error of its rung. We measure the latter rather than model it:
$e(g)$ is the excess loss of the model with every block at $g$
(\S\ref{sec:alloc}). We allocate by solving
\begin{equation}
  \min_{\{g_b\}}\;\sum_{b\in\mathcal{B}} \mathrm{imp}_b\, e(g_b)
  \qquad\text{s.t.}\qquad
  \frac{1}{|\mathcal{B}|}\sum_{b\in\mathcal{B}} r(g_b) \;\le\; \bar r ,
  \label{eq:objective}
\end{equation}
where $\bar r$ is the target rate. Because $\mathrm{imp}$ has unit mean within
every tensor, the objective of each uniform allocation is proportional to its
measured $e(g)$; between them, \eqref{eq:objective} is an empirical surrogate
for the loss.

The main design constraint is that \emph{every quantity that determines $g_b$
must be available to the decoder without a per-block entry in the bitstream.}
This rules out importance scores that read a block's own weights or on
calibration data, since these would have to be transmitted for every block. 
Per-channel quantities are acceptable, which cost a few thousandths of a bit per weight
(\S\ref{sec:decode}). \seed{} computes them from the checkpoint, so its
importance, like SeedLM's codec, reads no data.

\subsection{Data-free block importance}
\label{sec:importance}

\paragraph{Input side.}
For $\mathbf{y}=\mathbf{W}\mathbf{x}$, a perturbation changes the output by
$\mathbb{E}\lVert\Delta\mathbf{W}\mathbf{x}\rVert^2\approx\sum_j
a_j\lVert\Delta\mathbf{W}_{:,j}\rVert^2$ with $a_j=\mathbb{E}[x_j^2]$.
This quantity is estimated by AWQ, OPTQ and diagonal-Fisher methods from calibration data. In a
pre-norm transformer, five of the seven projections
($\mathbf{W}_q,\mathbf{W}_k,\mathbf{W}_v,\mathbf{W}_{\mathrm{gate}},\mathbf{W}_{\mathrm{up}}$)
read $\mathbf{x}=\boldsymbol{\gamma}\odot\mathrm{RMSNorm}(\mathbf{h})$ with a
stored gain $\boldsymbol{\gamma}$, so, with the channels of the normalized
vector taken as isotropic,
\begin{equation}
  a_j \;\approx\; \gamma_j^2 ,
  \label{eq:gamma}
\end{equation}
read directly from the checkpoint. The other two projections read tensors
produced by projections covered by \eqref{eq:gamma} and the second moment
propagates one hop:
\begin{equation}
  a^{(o)} = \mathrm{repeat}_{\mathrm{kv}}\!\big((\mathbf{W}_v\odot\mathbf{W}_v)\,a^{(\mathrm{in})}\big),
  \qquad
  a^{(\mathrm{down})}_c = \mathbb{E}\big[\mathrm{silu}(g_c)^2\big]\,
  \big((\mathbf{W}_{\mathrm{up}}\odot\mathbf{W}_{\mathrm{up}})\,a^{(\mathrm{in})}\big)_c ,
  \label{eq:prop}
\end{equation}
with $\mathbb{E}[\mathrm{silu}(g_c)^2]$ evaluated by Gauss--Hermite quadrature
for $g_c\sim\mathcal{N}\big(0,((\mathbf{W}_{\mathrm{gate}}\odot\mathbf{W}_{\mathrm{gate}})\,a^{(\mathrm{in})})_c\big)$.
No step reads a token; Appendix~\ref{app:approx} lists the approximations.

\paragraph{Output side.}
The Fisher of a linear layer is $F_{rc}=\mathbb{E}[g_r^2x_c^2]\approx\rho_r a_c$
under the Kronecker-factored approximation~\citep{martens2015kfac}, where
$\rho_r=\mathbb{E}[g_r^2]$ is the scale of the output gradient that a dimension
$a$ cannot represent. We estimate $\rho$ by the same propagation run backward
from an isotropic gradient at the logits,
$\mathbb{E}[g_{\mathrm{in},c}^2]\approx\sum_r W_{rc}^2\,\mathbb{E}[g_{\mathrm{out},r}^2]$,
and apply it to the rows of $\mathbf{W}_o$ and $\mathbf{W}_{\mathrm{down}}$,
which read the residual-stream gradient directly; all other rows receive $\rho=1$.

\paragraph{Block importance.}
\label{sec:blocking}
Input channels are sorted by $a$ before blocking, so a block holds eight
channels of similar scale and the variation lies between blocks, where the
allocator can use it. With both factors normalized to unit mean within each
tensor,
\begin{equation}
  \mathrm{imp}_b \;=\; \bar a_b\;\rho_{r(b)},
  \qquad
  \bar a_b = \tfrac{1}{B}\textstyle\sum_{j\in b} a_j .
  \label{eq:imp}
\end{equation}
The normalization discards the estimate's between-tensor scale, which is
dominated by a depth-dependent artifact of the propagation; even a calibrated
between-tensor scale makes the allocation worse (Appendix~\ref{app:unitmean}).
The estimate is least reliable at its bottom, where small differences between
small numbers decide demotions, so we clip it at its $f$-quantile $q$,
$\mathrm{imp}_b\leftarrow\max(\mathrm{imp}_b,q)$, with $f=0.06$ chosen once on
Llama-2-7B (Figure~\ref{fig:ablation}b) and fixed for every model.

\subsection{Rate--distortion allocation}
\label{sec:alloc}

\paragraph{Damage curve and hull.}
For each rung $g$ we build the uniform configuration and record its excess
loss $e(g)$ over FP16 on WikiText-2: $|G|$ scalars per model that fix the
slopes of the trade-off between rungs, not the rung of any block (measuring
them on the training split instead changes $1.3\%$ of the rungs of
Llama-3-8B; Appendix~\ref{app:hull}). We keep the lower convex hull of
$\{(r(g),e(g))\}$: rungs $g_1,\dots,g_J$ with rates $r_1<\dots<r_J$ and slopes
$s_j=\big(e(g_j)-e(g_{j+1})\big)/(r_{j+1}-r_j)$, decreasing in $j$.

\paragraph{Lagrangian and ladder.}
In \eqref{eq:objective}, the blocks are coupled only through the rate
constraint. Introducing a multiplier $\lambda\ge 0$ gives the Lagrangian
\begin{equation}
  \mathcal{L}\big(\{g_b\},\lambda\big) \;=\;
  \sum_{b\in\mathcal{B}}\Big[\,\mathrm{imp}_b\,e(g_b)+\lambda\,r(g_b)\,\Big]
  \;-\;\lambda\,|\mathcal{B}|\,\bar r
  \label{eq:lagrangian}
\end{equation}
which separates over blocks. By Everett's
theorem~\citep{everett1963generalized} the per-block minimizer for a given
$\lambda$ solves \eqref{eq:objective} at its own total rate
(Appendix~\ref{app:everett}); bisection on $\lambda$ meets the budget, the
classical bit-allocation argument of \citet{shoham1988efficient}. On a convex
hull the per-block minimizer has a closed form: rung $j{+}1$ beats rung $j$
exactly when $\mathrm{imp}_b\,s_j\ge\lambda$. Hence
\begin{equation}
  g_b \;=\; g_{\,j(b)},\qquad
  j(b) \;=\; 1+\big|\{\,j:\ \lambda/s_j \le \mathrm{imp}_b\,\}\big| ,
  \label{eq:ladder}
\end{equation}
which amounts to a ladder applied to $\mathrm{imp}$, with one comparison per block. Blocks
that share a column range have identical importance, so the last threshold
adjusts a tied group at once; a deterministic pass demotes members of that group
in block order until the budget is met exactly.

\subsection{Outlier columns}
\label{sec:outliers}

Objective~\eqref{eq:objective} charges each block the average error of its
rung, which is incorrect for blocks that cannot be represented well at any
rung.
In every model we examined, a few columns contain a single weight that
exceeds the dynamic range of a shared-exponent block by two orders of
magnitude. These are the super-weight columns identified by
\citet{yu2024superweight}, which multiply the largest activation in the
network.
The codec reproduces the peak but leaks part of it onto the
near-zero block-mates, which the massive activation amplifies: on Llama-2-7B, one such column carries $0.020$ of uniform SeedLM's $0.067$ excess loss. We
select candidate columns by dynamic range
($\max_i|W_{ij}|/\operatorname{median}_i|W_{ij}|\ge10^{3}$) or activation scale
($a_j\ge10^{2}$ times the tensor mean), both read from the checkpoint and
store the first four in FP16 on every model, $3\times10^{-5}$ bits per weight
on Llama-2-7B (Appendix~\ref{app:outliers}). The dynamic-range rule alone
recovers the published super-weight coordinates of Llama-2-7B, Llama-2-13B and
Llama-3-8B from weights. 

\subsection{Decoding without per-block side information}
\label{sec:decode}

The decoder reproduces $\{g_b\}$ from (i) the RMSNorm gains, which give $a$ for
five projection types by \eqref{eq:gamma}; (ii) a per-tensor table with
$a^{(o)}$, $a^{(\mathrm{down})}$ and $\rho$ for the rows of $\mathbf{W}_o$ and
$\mathbf{W}_{\mathrm{down}}$, computed by the encoder from the FP16 weights;
and (iii) a global table with the hull, $\lambda$, $q$, the tie counts, the
tensor offsets and the outlier list. It computes \eqref{eq:imp}, applies the
floor and \eqref{eq:ladder}, learns each block's length, reads that many bits
and reconstructs the block with \eqref{eq:codec}; no block carries a header.
The per-tensor table has one entry per row or column, $0.003$-$0.004$ bits
per weight in single precision. 
Because no rung depends on a decoded weight, tensors can be decoded in any order, on demand and in the order of the forward pass. The encoder allocates in a single pass based exactly on what the decoder holds, while a corrupted seed affects only its own block.
\S\ref{sec:exp} are payload rates; the tables add at most $0.005$ bits per
weight (Appendix~\ref{app:decode}).
\section{Experimental Results}
\label{sec:exp}

\subsection{Performance Analysis}
\label{sec:perf}

We evaluate \seed{} on Llama-2-7B, Llama-2-13B, Llama-3-8B and Mistral-7B by
WikiText-2 perplexity (2048-token windows) and zero-shot accuracy on five tasks
(LM Evaluation Harness v0.4.3), following \citet{shafipour2025seedlm}. Due to the absence of an open-source implementation, we present results for SeedLM through the faithful implementation built with the same codec and harness as
\seed{}; it matches the paper's weight-space results but has a higher
perplexity. S-Quant and AWQ~\citep{lin2024awq} are as published
\citep{wang2026squant,shafipour2025seedlm}. \seed{} includes its four FP16
outlier columns (\S\ref{sec:outliers}); rates are payload rates and the side
tables add at most $0.005$ bits per weight (Appendix~\ref{app:decode}). The
two rows of \seed{} differ only in the target rate $\bar r$ and the per-rung
compressions that dominate the encoding cost serve both
(Appendix~\ref{sec:cost}).

\begin{figure}[!htbp]
\centering
\includegraphics[width=0.333\linewidth]{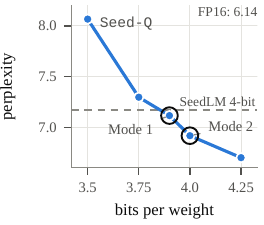}\hfill
\includegraphics[width=0.333\linewidth]{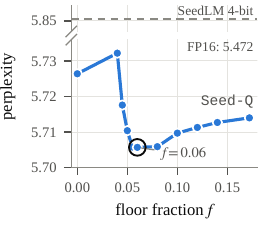}\hfill
\includegraphics[width=0.333\linewidth]{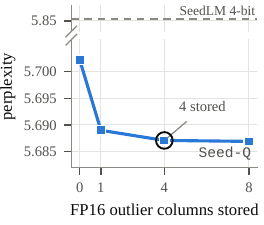}
\caption{Ablation of \seed{} (WikiText-2 perplexity). (a)~Rate, with the four
outlier columns; dashed: SeedLM at $4.0$ bits per weight; rings: Mode~1 and~2
(Table~\ref{tab:accuracy}). (b)~Floor $f$ (\S\ref{sec:importance}) at $4.0$
bits, column factor only, no outlier columns. (c)~Outlier columns at $4.0$ bits.}
\label{fig:ablation}
\end{figure}

\paragraph{Mode~1:}
Mode~1 is the lowest measured rate at which \seed{} reaches the perplexity of
SeedLM at $4.0$ bits per weight: $3.77$, $3.75$, $3.88$ and $3.78$
bits (Tables~\ref{tab:ppl} and~\ref{tab:accuracy}), $3$-$6\%$ fewer bits
and $0.40$\,GB of weight storage on Llama-2-13B; the crossing itself can lie
lower, near $3.85$ bits on Llama-3-8B (Figure~\ref{fig:ablation}a). 
The saving comes from treating the rate as a
budget: \seed{} can target any rate between the rungs of its grid, whereas
SeedLM's nearest configuration below $4.0$ bits, $(14,3)$ at $3.75$, has
$1.5\times$ the excess loss of its 4-bit point on Llama-2-13B.

\paragraph{Mode~2:}
At $4.0$ bits/weight \seed{} lowers SeedLM's perplexity from $5.85$ to
$5.68$, $5.18$ to $5.06$, $7.17$ to $6.92$ and $6.32$ to $5.96$, removing
$42\%$, $38\%$, $23\%$ and $32\%$ of its excess loss over FP16 and is at or
below the published SeedLM and AWQ perplexities on the three Llama models.
S-Quant's rate omits the per-block coefficient count that its decoder must read,
$0.19$ bits per weight with a fixed-length code, which places its $3.8$ bits
near $3.99$; in our implementation, its reconstruction error at $3.8$ bits is
$2.1$-$2.3\times$ SeedLM's at $4.0$ (Appendix~\ref{app:squant}).

\begin{table}[!ht]
\centering
\caption{WikiText-2 perplexities for LLaMA 2, LLaMA 3 and Mistral 4-bit weight representations evaluated on 2048-token contexts. \nr: Not Reported}
\label{tab:ppl}
\small
\begin{tabular}{@{}lccccc@{}}
\toprule
    \multirow{2}{*}{{Method}} & \multicolumn{4}{c}{{Model Perplexity (Bits/Weight)}} & {Mean} \\ \cline{2-5}
    {} & {LLaMA-2-7B}& {LLaMA-2-13B} & {LLaMA-3-8B} & {Mistral-7B} & {Bits}\dg \\
\midrule
    Baseline                                & 5.5 (16) & 4.9 (16) & 6.1 (16) & 5.25 (16) & 16 \\
    \cdashline{1-6}
\rowcolor{shade}
    {\seed~(Mode 1)*}                 & 5.85 (3.77) & 5.18 (3.75) & 7.17 (3.88) & 6.32 (3.78) & 3.79 \\
    S-Quant                                 & 5.7 (3.8)  & 5.0 (3.8) & 6.8 (3.8) & {\nr} & 3.8 \\
\rowcolor{shade}
    {\seed~(Mode 2)\bstars}           & 5.68 (4) & 5.06 (4) & 6.92 (4) & 5.96 (4) & 4 \\
    SeedLM\ddg                     & {5.85 (4)} & {5.18 (4)} & {7.17 (4)} & {6.32 (4)} & 4 \\
    AWQ                                     & 5.8 (4)  & 5.1 (4) & 7.1 (4) & {\nr} & {4}\\
\bottomrule
\end{tabular}
\begin{minipage}{0.97\linewidth}
\footnotesize
    {*} Mode 1: Lowest storage configuration. \bstars~Mode 2: Best PPL configuration for 4-bit. \dg~Excluding \nr. \\
    \ddg~Actual results obtained from faithful replication of SeedLM.
\end{minipage}
\end{table}

\paragraph{Zero-shot accuracy.}
SeedLM is $1.2$-$2.0$ points of mean accuracy below FP16 on the
Llama models and $3.8$ on Mistral-7B (Table~\ref{tab:accuracy}). Mode~2 raises
the mean accuracy by $0.5$-$1.1$ points on all four models, recovering
$27$-$55\%$ of the loss and Mode~1 stays within about one point of SeedLM,
the scale of single-task sampling error (about $0.9$ points on ARC-Easy).

\begin{table}[!h]
\caption{Performance comparison across different models and zero-shot tasks for around 4-bit configurations. Entries that ran out of memory in our setup are marked with OOM. \nr: Not Reported.}
\label{tab:accuracy}
\centering
\resizebox{1.0\columnwidth}{!}{%
\begin{tabular}{llccccccc}
    \hline
    \multirow{2}{*}{{Model}} & \multirow{2}{*}{{Method}} & \multicolumn{6}{c}{{Zero-Shot Task Accuracy (\%)}} \\ \cline{3-9}
     {} & {} & {Bits} & {ARC-Easy} & {ARC-Challenge} & {HellaSwag} & {WinoGrande} & {BoolQ} & {Mean} \\
    \hline
    \multirow{6}{*}{{LLaMA-2-7B}}
        & Baseline  & 16  & 74.58 & 46.33 & 75.98 & 69.06 & 77.74 & 68.74 \\
        \cdashline{2-9}
        & \ccbb {\seed~(Mode 1)*}      & \ccbb {3.77} & \ccbb {71.97} & \ccbb {43.43} & \ccbb {73.60} & \ccbb {68.03} & \ccbb {74.98} & \ccbb {66.40} \\
        & S-Quant   & 3.8 & 73.36 & 44.55 & 74.51 & 68.47 & 77.34 & 67.65 \\
        & \ccbb {\seed~(Mode 2)\bstars}  & \ccbb {4}    & \ccbb {72.31} & \ccbb {44.45} & \ccbb {74.69} & \ccbb {69.53} & \ccbb {75.90} & \ccbb {67.38} \\
        & SeedLM$^\ddagger$      & 4   & 72.39 & 43.60 & 73.79 & 67.72 & 76.85 & 66.87 \\
        & AWQ       & 4   & 70.58 & 43.94 & 74.96 & 68.75 & 78.29 & 67.30 \\
    \hline
    \multirow{6}{*}{Llama 2 13B}
        & Baseline  & 16  & 77.44 & 48.98 & 79.38 & 72.22 & 80.55 & 71.71 \\
        \cdashline{2-9}
        & \ccbb {\seed~(Mode 1)*}     & \ccbb {3.75} & \ccbb {77.23} & \ccbb {50.34} & \ccbb {77.73} & \ccbb {71.74} & \ccbb {80.52} & \ccbb {71.51} \\
        & S-Quant   & 3.8 & 77.02 & 49.93 & 78.55 & 72.81 & 79.33 & 71.53 \\
        & \ccbb {\seed~(Mode 2)\bstars} & \ccbb {4} & \ccbb {76.85} &\ccbb  {48.55} &\ccbb  {78.03} &\ccbb  {72.45} &\ccbb  {79.88} &\ccbb  {71.15} \\
        & SeedLM$^\ddagger$     & 4   & 76.73 & 48.21 & 77.23 & 71.98 & 78.23 & 70.47 \\
        & AWQ       & 4   & 77.44 & 49.32 & 78.57 & 71.90 & 78.47 & 71.14 \\
    \hline
    \multirow{6}{*}{{LLaMA-3-8B}}
        & Baseline  & 16  & 76.81 & 52.73 & 76.97 & 72.93 & 81.87 & 72.26 \\
        \cdashline{2-9}
        & \ccbb {\seed~(Mode 1)*}     & \ccbb {3.88} & \ccbb {76.68} & \ccbb {49.15} & \ccbb {76.70} & \ccbb {72.61} & \ccbb {70.98} & \ccbb {69.22} \\
        & S-Quant   & 3.8 & 76.68 & 49.89 & 76.72 & 73.12 & 80.84 & 71.45 \\
        & \ccbb {\seed~(Mode 2)\bstars} & \ccbb {4}    & \ccbb {76.26} & \ccbb {50.17} & \ccbb {76.26} & \ccbb {72.69} & \ccbb {78.69} & \ccbb {71.04} \\
        & SeedLM$^\ddagger$     & 4   & 77.31 & 50.00 & 76.80 & 73.01 & 74.04 & 70.23 \\
        & AWQ       & 4   & 74.49 & 51.54 & 78.03 & 73.09 & 80.40 & 71.51 \\
    \hline
    \multirow{4}{*}{{Mistral-7B}}
        & Baseline  & 16  & 79.55 & 53.92 & 81.06 & 74.03 & 83.64 & 74.44 \\
        \cdashline{2-9}
        & \ccbb {\seed~(Mode 1)*}     & \ccbb {3.78} & \ccbb {75.55} & \ccbb {47.70} & \ccbb {76.88} & \ccbb{72.38} & \ccbb {80.70} & \ccbb {70.64} \\
        & S-Quant   & {\nr}   & {\nr} & {\nr} & {\nr} & {\nr} & {\nr} & {\nr} \\
        & \ccbb {\seed~(Mode 2)\bstars} & \ccbb {4}    & \ccbb {76.43} & \ccbb {50.77} & \ccbb {77.49} & \ccbb {71.98} & \ccbb {81.99} & \ccbb {71.73} \\
        & SeedLM$^\ddagger$     & 4   & 75.13 & 48.55 & 76.79 & 70.80 & 81.90 & 70.63 \\
        & AWQ                            & 4   & {\nr} & {\nr} & {\nr} & {\nr} & {\nr} & {\nr} \\
    \hline
\end{tabular}
}
\footnotesize \raggedright \\
    {*} Mode 1: Lowest storage configuration. \bstars~Mode 2: Best PPL configuration for 4-bit. \\
    \ddg~Actual results obtained from faithful replication of SeedLM. 
\end{table}

\paragraph{Ablation.}
Figure~\ref{fig:ablation} varies one factor of \seed{} at a time. (a)~Its
perplexity falls steeply up to $3.75$ bits per weight and flattens beyond
$4.0$. (b)~The floor raises the share of SeedLM's
excess loss removed from $32\%$ to $37.5\%$, remaining flat for $f\in[0.055,0.08]$;
$f=0.06$ is used for every model. (c)~Four outlier columns remove $6\%$ of
\seed{}'s excess loss because the allocator already spends higher rungs on
the same columns; more columns change nothing.

\subsection{Overhead Analysis}
\label{sec:hw}

We evaluate weight-reconstruction kernels for SeedLM, S-Quant and \seed~using Yosys with the Nangate 45 nm standard-cell library. Timing and power are estimated with OpenSTA at the typical corner (1.10 V, 25°C). We report mapped cell area, estimated total cell power and register-to-register path delay. Power is evaluated at 100 MHz using 128 matched synthetic blocks. All kernels produce Q16.16 weights and are evaluated with 16-bit seeds and (k=3), requiring a 561-cycle interval per eight-weight block. The details of hardware implementation is described in Appendix~\ref{app:seedq_hw}.

Please note that we independently implement SeedLM and S-Quant based on their published weight-reconstruction formulations since no open-source implementations are available. To enable a fair comparison, we use a common output precision, technology library, clock frequency and synthesis and power-estimation flow.

\noindent
\begin{minipage}[t]{0.37\linewidth}
\raggedright
\vspace{0pt}
Table \ref{tab:overhead} shows that \seed~requires 52.6\% less cell area
and 39.6\% less estimated power than the implemented S-Quant variant,
with a 42.8\% shorter register path. Relative to SeedLM with fixed
configurations, \seed~incurs 30.8\% area and 30.5\% power overhead while
supporting runtime seed widths of 8-16 bits and basis counts of 2-6.
Each implementation is evaluated at a 16-bit seed and $k=3$, with 561 cycles per eight-weight block.
\end{minipage}%
\hfill
\begin{minipage}[t]{0.6\linewidth}
\vspace{0pt}
\centering
\small
\resizebox{0.99\linewidth}{!}{%
\begin{tabular}{@{}lrrr@{}}
\toprule
Component & Area ($\mu$m$^2$) & Power (mW) & Reg. path (ns) \\
\midrule
\multicolumn{4}{@{}l}{\textbf{\underline{PE (common)}}} \\
Single PE & 5,188.1 & 0.978 & 2.094 \\
$16\times16$ PE array & 1,328,153.6 & 102.346 & 2.383 \\
\midrule
\multicolumn{4}{@{}l}{\underline{\textbf{Weight generator}}} \\
SeedLM & 2,891.7 & 0.211 & 1.671 \\
S-Quant$^{\dagger}$ & 7,970.4 & 0.456 & 3.200 \\
\rowcolor{shade}
\seed{} & 3,781.7 & 0.275 & 1.831 \\
\bottomrule
\end{tabular}
}
\captionof{table}{Hardware cost of the shared PE design and weight generators.
Area is mapped cell area; power and register-path delay are pre-layout
estimates. The array contains 256 instances of the common PE design.\\
$^{\dagger}$S-Quant uses a direct-LFSR architecture.}
\label{tab:overhead}
\end{minipage}

At an effective representation rate of 3.77 bits per weight, \seed~reduces ideal weight-transfer volume by 76.4\% relative to FP16, corresponding to a theoretical 4.24x increase in bandwidth-limited weight-delivery throughput.

\subsection{Security Analysis}
\label{sec:security}

Targeted bit flips can degrade neural-network predictions by corrupting a small number of stored parameters~\citep{rakin2019bitflip,yao2020deephammer}. SeedLM~\citep{shafipour2025seedlm} and S-Quant~\citep{wang2026squant} evaluate compression and efficiency, but do not report bit-flip security evaluations. We measure how selected faults change reconstructed weights and model performance. 

\subsubsection{Fault model and propagation}

We consider a single bit flipped after encoding and before decoding, in a weight, scale, seed, or allocation input. We measure its effect on reconstructed weights and inference quality. This is a software fault study, without physical memory injection or a search for an attack that induces a chosen prediction. 

Each scheme is compared with its own clean reconstruction, excluding ordinary quantization error from these measurements. We retain $\mathbf W\in\mathbb R^{R\times C}$ and block size $B$ from Section~\ref{sec:codec}. Let $\widehat{\mathbf W}$ and $\widetilde{\mathbf W}$ be the clean and corrupted reconstructions. For this comparison, the perturbation in Section~\ref{sec:problem} is $\Delta\mathbf W=\widetilde{\mathbf W}-\widehat{\mathbf W}$. We report the number of changed weights~($N_{\mathrm{flip}}$), their fraction~($p_{\mathrm{flip}}$) and relative Frobenius error~($\eta$):
\begin{equation}
 N_{\mathrm{flip}}=\sum_{r=1}^{R}\sum_{c=1}^{C}
   \mathbf{1}[\Delta W_{rc}\ne0],\qquad
 p_{\mathrm{flip}}=\frac{N_{\mathrm{flip}}}{RC},\qquad
 \eta=\frac{\|\Delta\mathbf W\|_F}{\|\widehat{\mathbf W}\|_F}.
 \label{eq:security_metrics}
\end{equation}

\par
\noindent
\begin{minipage}[t]{.33\linewidth}
\raggedright
Figure~\ref{fig:footprint} reports the number of changed weights:
1 for No-Q, 70 for AWQ and 8 for SeedLM. Under simulated propagation,
S-Quant and \seed{} affect 7.31\,M and 16.77\,M weights,
respectively. The first three footprints follow direct bit injection;
the latter two follow simulated substitution of later blocks. 
The relative errors $\eta$ are $9.70\times10^{-4}$, 0.88 and 1.41 for SeedLM, S-Quant and \seed, respectively. 
The experimental setup used for security evaluation is described in Appendix~\ref{app:sec_setup}.

\end{minipage}\hfill
\begin{minipage}[t]{.65\linewidth}
\captionsetup{type=figure,font=small,skip=4pt}
\centering
\raisebox{\dimexpr\ht\strutbox-\height\relax}{%
\includegraphics[width=.85\linewidth]{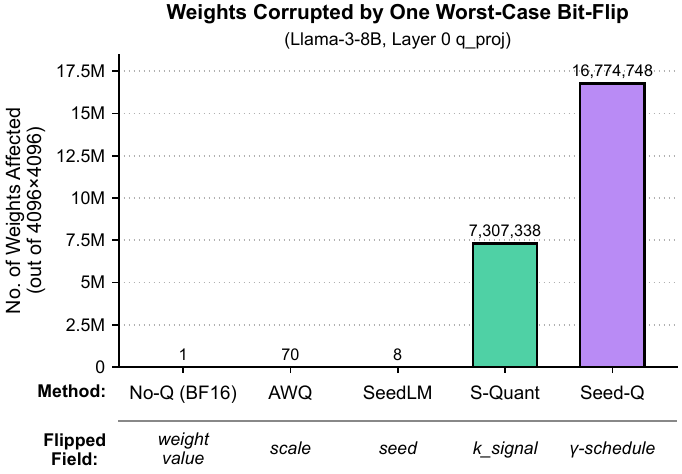}}
\caption{Affected weights in the Llama-3-8B query projection:
43.56\% for S-Quant and 99.99\% for \seed. No-Q (BF16),
AWQ and SeedLM each affect $<0.001\%$.}
\label{fig:footprint}
\end{minipage}
\par

\begin{figure}[t]
\centering
\begin{subfigure}[t]{.49\linewidth}
\centering
\includegraphics[width=0.95\linewidth]{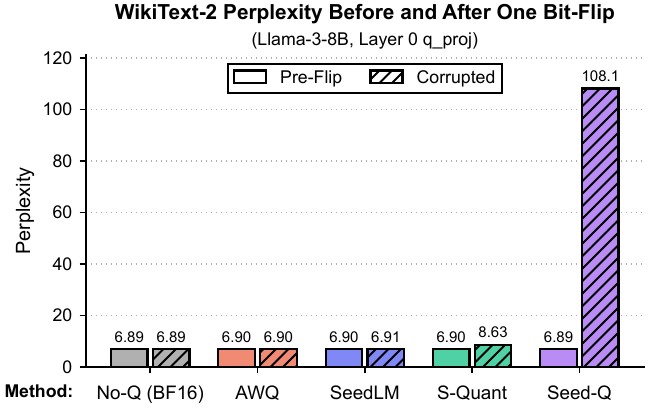}
\caption{WikiText-2 perplexity.}
\label{fig:ppl}
\end{subfigure}\hfill
\begin{subfigure}[t]{.49\linewidth}
\centering
\includegraphics[width=0.95\linewidth]{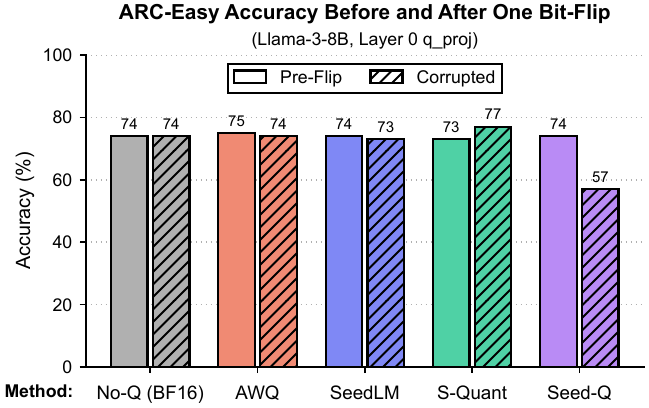}
\caption{ARC-Easy accuracy on 100 questions.}
\label{fig:acc}
\end{subfigure}
\caption{Model performance before and after the selected fault.
\emph{Pre-Flip} denotes each scheme's uncorrupted state.
Lower PPL and higher accuracy indicate better performance.
S-Quant and \seed{} use simulated propagation.}
\label{fig:performance}
\end{figure}

\subsubsection{Security implications of the results}

\paragraph{Fault amplification in PPL.}
Figure~\ref{fig:performance}(a) shows negligible PPL changes for No-Q
and AWQ: increases of 0.01\% and 0.06\%, computed before rounding.
Among the seed-based schemes, SeedLM changes from 6.90 to 6.91,
S-Quant from 6.90 to 8.63 ($+25.09\%$) and \seed{} from
6.89 to 108.11 ($+1{,}468.43\%$). The larger response for
\seed{} depends in part on its earlier allocation mismatch.
so this comparison does not establish a ranking across fault positions.

\paragraph{Fault amplification in accuracy.}
Figure~\ref{fig:performance}(b) shows unchanged No-Q accuracy at 74\%
and an AWQ decrease from 75\% to 74\%. Among the seed-based schemes,
SeedLM falls from 74\% to 73\%, S-Quant rises from 73\% to 77\% and
\seed{} falls from 74\% to 57\%, a 17-percentage-point drop.
S-Quant's gain does not establish robustness, since its PPL worsens and
each answer in this 100-question sample changes accuracy by 1 percentage
point. 
not claim statistical significance.

\paragraph{Detection and countermeasures.}
The allocation-dependent amplification in \seed{} could make a targeted fault visible to a monitor that compares loss on trusted inputs with a clean reference. Such a check would need to reject corrupted state before its outputs are used and reload a trusted copy. However, fault amplification alone does not provide bit-flip attack resistance: without a working check and recovery path, the same fault can cause severe service degradation. 

Robust protection against targeted bit-flip attacks should cover the compressed payload and every input that determines its interpretation. Error-correcting memory handles faults within its correction capability, but does not eliminate targeted Rowhammer attacks~\citep{kamadan2025eccfail}. Authentication tags or cryptographic hashes checked against protected references can verify seeds, coefficients, scales and allocation metadata before use. Runtime weight signature checks provide another option~\citep{li2021radar}. For
\seed{} protections must include $\boldsymbol{\gamma}$ and the
global decoder table: $G$, $s_j$, $\lambda$, $q$ and the tie count
(Section~\ref{sec:decode}). Such countermeasures require separate comprehensive evaluation to determine their coverage and overheads.

\section{Conclusion}
We have presented \seed, a seed-based model-weight generation scheme for LLMs that delivers improved compression performance compared with existing seed-based solutions while simultaneously providing quantifiable protection against bit-flip attacks. \seed~exploits the intrinsic skew in model-weight sensitivity to preferentially allocate seed budgets within an LFSR-based compression framework. By eliminating the need for allocation metadata, \seed~reduces storage overhead while enabling weights to be reproducibly decoded using simple, low-cost hardware, as observed in our custom hardware implementation. We further show that \seed~protects critical weight bits against bit-flip attacks by greatly amplifying the impact of malicious bit modifications on model behavior, thereby making such attacks readily detectable. Finally, our analysis shows that \seed~is scalable and flexible, supporting diverse model architectures and sizes.


\bibliography{conference}
\bibliographystyle{conference}

\newpage

\appendix

\section*{Appendix}
\section{Method details}
\label{app:method}

\subsection{Approximations behind the importance}
\label{app:approx}

\paragraph{Input side.}
Equation~\eqref{eq:gamma} writes $a_j=\gamma_j^2\,\mathbb{E}[n_j^2]$ with
$\mathbf{n}=\mathrm{RMSNorm}(\mathbf{h})$ and takes $\mathbb{E}[n_j^2]\approx1$.
RMSNorm fixes the mean square of $\mathbf{n}$ over its channels to one for
every token; it does not fix the second moment of each channel, so the
approximation assumes the channels are isotropic. Equation~\eqref{eq:prop}
follows the value path only: attention averages value vectors over positions
with weights shared by all channels of a head, which scales each head's
channels by a head-dependent factor that the $a^{(o)}$ term omits. The
$a^{(\mathrm{down})}$ term treats the gate and up branches as independent. All
three terms assume a diagonal $\mathbb{E}[\mathbf{x}\mathbf{x}^{\!\top}]$, as
diagonal-Fisher methods do.

\paragraph{Output side.}
The multiplicative form $F_{rc}\approx\rho_r a_c$ follows from the
Kronecker-factored approximation
$\mathbb{E}[g_r^2x_c^2]\approx\mathbb{E}[g_r^2]\,\mathbb{E}[x_c^2]$. The
backward proxy starts from an isotropic gradient at the logits, the only
data-free choice; one hop through the language-model head gives a per-channel
gradient scale on the residual stream and each layer, taken in reverse order,
maps it back through its own matrices. It keeps what the weights determine and
drops every factor that needs activations: the derivative of SiLU and the
other branch's activation in the MLP, the attention weights and the
Jacobians of the normalizations. The rows of $\mathbf{W}_o$ and
$\mathbf{W}_{\mathrm{down}}$ write to the residual stream and read its
gradient directly. The rows of $\mathbf{W}_v$, $\mathbf{W}_{\mathrm{gate}}$
and $\mathbf{W}_{\mathrm{up}}$ receive it only through a second matrix and
through the attention or SiLU the proxy drops; propagating $\rho$ to them
anyway lowers the damage on Llama-2-7B from $0.0412$ to $0.0395$ and we report
the conservative variant on every model. The rows of $\mathbf{W}_q$ and
$\mathbf{W}_k$ reach the loss only through the softmax Jacobian, for which we
have no derivation: in a fixed-budget allocation a factor without a derivation
does not merely fail to help, it moves bits away from the blocks that need
them.

\subsection{Normalization within tensors}
\label{app:unitmean}

The propagation in \eqref{eq:prop} accumulates a multiplicative, depth-dependent artefact across layers that carries no information about importance and even a calibrated between-tensor scale hurts when handed to the allocator. On Llama-2-7B, adding the per-tensor mean of a calibrated Fisher to the within-tensor ranking makes the allocation $5\%$ worse at the same floor ($0.0428\to0.0450$), while on Llama-3-8B the same Fisher normalized to unit mean per tensor, with no between-tensor information at all, still removes $27\%$ of SeedLM's excess loss. \seed{} therefore ranks blocks within tensors only. Every tensor keeps unit mean importance, so the share of the budget a tensor receives is set by the shape of its importance distribution defined by how many of its blocks clear the thresholds of \eqref{eq:ladder}) and not by its scale.

\subsection{The floor}
\label{app:floor}

With the column factor alone at $4.0$ bits per weight on Llama-2-7B, the
perplexity is flat to within $0.0002$ for $f\in[0.055,0.08]$ and the damage
within $2\%$ for $f\in[0.05,0.10]$ (Figure~\ref{fig:ablation}b). The effect
does not depend on the column factor: with the row factor added it lowers the
damage from $0.0428$ to $0.0395$. The value $q$ the floor produces is
model-specific and travels in the global table, so the decoder never computes
a quantile. Equation~\eqref{eq:objective} and the ladder use the floored
importance.

\subsection{The damage curve}
\label{app:hull}

$e(g)$ is measured in loss rather than weight space because the loss is
markedly superlinear in reconstruction error at these rates, which changes the
hull's slopes; a weight-space $e(g)$ would make \seed{} entirely data-free. The
rung damages were measured on the WikiText-2 text on which
Section~\ref{sec:exp} reports perplexity. Measured on the training split
instead, the damages of Llama-3-8B move by $0$-$3.4\%$, the rung of $1.3\%$
of the blocks changes at $4.0$ bits per weight and the test perplexity moves
from $7.000$ to $7.005$ ($6.916\to6.918$ with the outlier columns). A single
curve serves the whole model because a rung's error depends on the rung and
hardly on the block: the error curves of the seven projection types agree to
within a few percent. The assumption fails for blocks the format cannot
represent at any rung, which \S\ref{sec:outliers} handles outside the hull. On
the models of Section~\ref{sec:exp} the hull keeps four to eight rungs.

\subsection{Everett's theorem and ties}
\label{app:everett}

For fixed $\lambda$ the minimiser of \eqref{eq:lagrangian} is obtained by
minimising $\mathrm{imp}_b\,e(g)+\lambda\,r(g)$ separately for every block. By
Everett's theorem, an allocation $\{g_b\}$ that minimises
$\mathcal{L}(\cdot,\lambda)$ minimises $\sum_b\mathrm{imp}_b\,e(g_b)$ over all
allocations whose total rate does not exceed its own,
$R^\star(\lambda)=\sum_b r(g_b)$: for any $\{g'_b\}$ with
$\sum_b r(g'_b)\le R^\star$,
\[
\sum_b \mathrm{imp}_b e(g_b)+\lambda R^\star
 \le \sum_b \mathrm{imp}_b e(g'_b)+\lambda\sum_b r(g'_b)
 \le \sum_b \mathrm{imp}_b e(g'_b)+\lambda R^\star .
\]
$R^\star(\lambda)$ is non-increasing in $\lambda$, so bisection drives it to
the budget; rates are discrete, so the budget can fall between two attainable
values and the tie pass closes that gap by demoting blocks that sit exactly at
the threshold, each trading damage for rate at price $\lambda$. The result is
optimal for \eqref{eq:objective} up to one block's rate step. Monotonicity of
the ladder follows from the decreasing slopes of the hull. Under the objective
the allocation cannot be worse than the uniform one, which is feasible;
whether the measured loss follows the objective is a separate question
(Section~\ref{sec:exp}). The implementation also confines a block to rungs
within $1.0$ bit per weight below and $1.5$ above the uniform rung; no hull
rung on our grids lies outside that band, so it never binds.

How many blocks move depends on the dispersion of the importance and on the
hull. On Llama-2-7B, where the data-free importance is much less dispersed than the true sensitivity, most blocks stay at the uniform rung. The budget moves to the few percent of blocks the estimator identifies with confidence: the high-activation channels of the residual stream and the MLP. A comparable number of demotions pays for these promotions.

\subsection{Outlier columns}
\label{app:outliers}

A shared-exponent block format has a dynamic range of a few octaves. On a
super-weight block the codec reproduces the peak and leaks a fraction of it
onto the seven near-zero neighbours, each amplified by the massive activation;
on Llama-2-7B this single column carries $0.020$ of the $0.067$ excess loss of
uniform 4-bit SeedLM and $0.008$ of the $0.055$ of round-to-nearest
quantization with groups of 128. No rung of a block format represents such a
block, so \seed{} gives these columns an escape mode outside $G$: the whole column
is stored in FP16 and overwrites the codec's reconstruction at the decoder.

A column $j$ is a candidate if
$\max_i|W_{ij}|/(\operatorname{median}_i|W_{ij}|+\epsilon)\ge10^3$ or
$a_j\ge10^2$ times the tensor mean; each rule keeps at most its 32 largest
values, ties broken by tensor and column index. Each list is sorted by its own
statistic, the candidates are ranked by their better position in the two
lists and the first four are stored on every model; the number was set once
on Llama-2-7B, where most of what the list removes is removed by its first
four columns (Figure~\ref{fig:ablation}c). The activation rule covers models
whose massive-activation channels carry no weight spike. The cost is counted
conservatively, as the FP16 values plus the block bits they replace plus 24
bits of tensor and column index per column: $3\times10^{-5}$ bits per weight on
Llama-2-7B and below $5\times10^{-4}$ on every model tested. The columns are
stored at the precision of the reference model, FP16, without seed
compression; for checkpoints released in BF16 the reference is the FP16 cast.

\subsection{Encoding and decoding}
\label{app:decode}

Algorithms~\ref{alg:encode} and~\ref{alg:decode} are the encoder and the
decoder. Nothing in the decoder reads a reconstructed weight: block lengths depend only on the RMSNorm gains and the
two tables, so tensors can be decoded in any order, blocks are parsed left to
right inside a tensor and the per-tensor offsets give random access between
tensors. A fault in a block's seed or coefficients corrupts that block only; a
fault in either table changes how a whole tensor is parsed and a checksum over
the tables, whose size is negligible, detects it.

\paragraph{Agreement.}
Encoder and decoder evaluate \eqref{eq:imp}, the floor and the ladder on the
same stored numbers, so their allocations agree if their arithmetic does: the
factors are stored in single precision and evaluated in double precision, in a
fixed order for the block means and the unit-mean normalization, with ties
broken by block index. A fixed-point decoder would instead fix its own formats
for the stored factors, the thresholds $\lambda/s_j$ and the products and the
encoder would allocate in those formats.

\paragraph{Rates.}
The payload rate $|\mathcal{B}|^{-1}\sum_b r(g_b)$ is held at $\bar r$. The
effective rate adds the per-tensor table (single precision: $0.0037$,
$0.0029$, $0.0039$ and $0.0039$ bits per weight on Llama-2-7B, Llama-2-13B,
Llama-3-8B and Mistral-7B), the outlier columns and their indices and the global table ($O(|G|)$ numbers plus two per tensor),
at most $0.005$ bits per weight in all. SeedLM's effective rate equals its
payload rate.

\paragraph{Blocking.}
The column permutation carries no per-block metadata: $\sum_j W_{ij}x_j$ does
not depend on the order of the sum (up to floating-point rounding) and the
decoder recomputes the order from $a$. It is not free in hardware: the decoder
writes each reconstructed row through the inverse permutation, or the
activations are read in permuted order, which costs address generation or a
row buffer.

\begin{algorithm}[h]
\caption{\seed{} encoder. The decoder (Algorithm~\ref{alg:decode}) recomputes line~3 from the gains and $P$ and applies the stored $q$, $\lambda$ and tie counts.}
\label{alg:encode}
\begin{algorithmic}[1]
\Require FP16 weights $\{\mathbf{W}^{(t)}\}$, RMSNorm gains $\{\boldsymbol{\gamma}\}$;
  rung damages $e(g)$; target rate $\bar r$; floor fraction $f$
\State $H\leftarrow$ lower convex hull of $\{(r(g),e(g))\}$, slopes $s_j$
  \Comment{\S\ref{sec:alloc}}
\State $P\leftarrow\{a^{(o)},a^{(\mathrm{down})},\rho^{(o)},\rho^{(\mathrm{down})}\}$
  for every layer, from the FP16 weights \Comment{\eqref{eq:prop}, \S\ref{sec:importance}}
\State $\mathrm{imp}_b\leftarrow\bar a_b\,\rho_{r(b)}$ for every block, $a$ from
  $\boldsymbol{\gamma}^2$ or $P$, unit mean per tensor \Comment{\eqref{eq:imp}}
\State $q\leftarrow$ $f$-quantile of $\{\mathrm{imp}_b\}$;\
  $\mathrm{imp}_b\leftarrow\max(\mathrm{imp}_b,q)$
\State $\lambda\leftarrow$ bisection to mean rate $\bar r$ under \eqref{eq:ladder};
  demote the last tied group in block order to meet $\bar r$ exactly
\State $O\leftarrow$ the first four outlier columns \Comment{\S\ref{sec:outliers}}
\For{each tensor $t$ and each block $b$ of $\mathbf{W}^{(t)}$}
  \State emit the seed and coefficients of $\mathbf{w}_b$ at rung $g_{j(b)}$:
    $\ell(g_{j(b)})$ bits, no header \Comment{\eqref{eq:codec}}
\EndFor
\State emit the FP16 columns $O$, the table $P$ and $(H,\lambda,q,\text{tie counts},\text{offsets},O)$
\end{algorithmic}
\end{algorithm}

\begin{algorithm}[h]
\caption{\seed{} decoder}
\label{alg:decode}
\begin{algorithmic}[1]
\Require bitstream; RMSNorm gains $\{\boldsymbol{\gamma}\}$; per-tensor table $P$;
  global table $(H,\lambda,q,\{T_t\},\{\mathrm{offset}_t\},O)$
\For{each tensor $t$, in any order, from $\mathrm{offset}_t$}
  \State $\mathrm{imp}_b\leftarrow\bar a_b\,\rho_{r(b)}$ from $\boldsymbol{\gamma}^2$ and $P$,
    unit mean over $t$;\ $\mathrm{imp}_b\leftarrow\max(\mathrm{imp}_b,q)$
  \For{each block $b$ in order}
    \State $g_b\leftarrow g_{j(b)}$ by \eqref{eq:ladder}, the first $T_t$ tied blocks
      taking the lower rung \Comment{same rule, same inputs}
    \State read $\ell(g_b)$ bits;\ $\hat{\mathbf{w}}_b\leftarrow\mathbf{U}(s)\hat{\mathbf{t}}$
  \EndFor
  \State apply the inverse column permutation; overwrite the FP16 columns of $t$ in $O$
\EndFor
\end{algorithmic}
\end{algorithm}

\subsection{Cost}
\label{sec:cost}

The encoder's dominant cost is SeedLM's exhaustive seed search, $2^{S}$
least-squares fits per block, run once for every rung of the grid: the uniform
builds that measure $e(g)$ are SeedLM compressions at different rungs, a rung
with $S=14$ costing a quarter of one with $S=16$ and a rung with $S\le12$ less
than a sixteenth. The grid therefore costs a small multiple of SeedLM's single
compression (on Llama-3-8B the $(16,3)$ rung alone took 6.6~hours on four
GPUs); it is paid once per model and reused for every target rate. The
importance, the outlier statistics and the allocation add one pass over the
checkpoint and one comparison per block.

At the decoder, the reconstruction primitive is the same as SeedLM's: one LFSR expansion and $k_b$ multiply--accumulates per weight. Its cost differs, however, because blocks promoted to $k_b>3$ do proportionally more work. At $4.0$ bits per weight, the mean $k_b$ is $3.02$, $3.06$ and $3.00$ on Llama-2-13B, Llama-3-8B and Mistral-7B, respectively, and below $3.13$ on Llama-2-7B. The decoder also evaluates \eqref{eq:imp} and the ladder using one product and at most $J$ comparisons per block, with the table held on chip. It then parses variable-length blocks and writes the outlier columns. Section~\ref{sec:hw} reports the reconstruction kernel with a configurable seed length and coefficient count; the ladder evaluation and inverse column permutation are not included in that measurement.

\section{Analysis details}
\label{app:analysis}

\subsection{Tail precision of the estimator}
\label{app:tail}

Table~\ref{tab:tail} scores \seed{}'s importance against the empirical diagonal
Fisher of Llama-2-7B (128 windows of 2048 tokens of WikiText-2 train, per
weight, averaged over each block on the allocator's own blocking). ``Mass''
is the share of the total Fisher over the model held by the blocks in the
named part of a ranking; ``exposure'' is the share of those blocks that lie in
the Fisher's top decile (a random pick gives $10\%$ globally and each type's
own share of the top decile per type).

\begin{table}[h]
\centering
\small
\caption{Tail precision on Llama-2-7B, $8.1\times10^{8}$ blocks. Global rows:
\seed{}'s ranking against the Fisher's own. Per-type rows report the same statistics within each projection type: the share of the type's Fisher mass in \seed{}'s
top $1\%$ of that type, the share held by \seed{}'s bottom $30\%$ against the
Fisher's own bottom $30\%$ and the exposure of \seed{}'s bottom $3\%$; ``rand''
is the type's share of the global top decile, the exposure a random pick would
have.}
\label{tab:tail}
\setlength{\tabcolsep}{4pt}
\begin{tabular}{lcccc}
\toprule
\multicolumn{5}{l}{\emph{Global}} \\
part of the ranking & \seed{} mass & Fisher mass & \seed{} exposure & random \\
\midrule
top $0.01\%$   & $42.5\%$ & $59.0\%$ & & \\
top $1\%$      & $59.3\%$ & $72.2\%$ & & \\
top $10\%$     & $65.6\%$ & $82.0\%$ & & \\
bottom $3\%$   & $6.34\%$ & $0.006\%$ & $17.3\%$ & $10\%$ \\
bottom $10\%$  & $12.4\%$ & $0.095\%$ & $21.0\%$ & $10\%$ \\
bottom $30\%$  & $19.4\%$ & $1.73\%$  & $14.3\%$ & $10\%$ \\
\midrule
\multicolumn{5}{l}{\emph{Per projection type}} \\
type & top-$1\%$ mass & \shortstack{bottom-$30\%$ mass\\(\seed{} / Fisher)} & bottom-$3\%$ exp. & rand \\
\midrule
$\mathbf{W}_q$              & $57\%$ & $0.24\%$ / $0.012\%$ & $0.0\%$  & $2.6\%$ \\
$\mathbf{W}_k$              & $45\%$ & $0.24\%$ / $0.011\%$ & $0.1\%$  & $2.3\%$ \\
$\mathbf{W}_v$              & $69\%$ & $7.6\%$ / $0.21\%$   & $0.1\%$  & $25.8\%$ \\
$\mathbf{W}_o$              & $54\%$ & $2.4\%$ / $0.30\%$   & $11.0\%$ & $41.8\%$ \\
$\mathbf{W}_{\mathrm{gate}}$& $9\%$  & $1.8\%$ / $0.60\%$   & $23.4\%$ & $2.4\%$ \\
$\mathbf{W}_{\mathrm{up}}$  & $5\%$  & $3.4\%$ / $0.80\%$   & $49.6\%$ & $6.0\%$ \\
$\mathbf{W}_{\mathrm{down}}$& $97\%$ & $2.6\%$ / $1.24\%$   & $12.0\%$ & $9.5\%$ \\
\bottomrule
\end{tabular}
\end{table}

The Fisher is concentrated far beyond a rank correlation's reach: its top $0.01\%$ of blocks holds $59\%$ of its mass, the top $1\%$ of $\mathbf{W}_{\mathrm{down}}$ alone $42\%$, the top $1\%$ of $\mathbf{W}_v$ $24\%$. \seed{}'s top decile captures $80\%$ of what the Fisher's own top decile holds at every budget, with a membership precision of only $16$-$24\%$: it misranks middling blocks into the top and finds the enormous ones. 
At the bottom the ranking of $\mathbf{W}_{\mathrm{gate}}$ and $\mathbf{W}_{\mathrm{up}}$ is inverted: their bottom $3\%$ is $10\times$ and $8\times$ more likely than a random block of the same type to be critical, because $a_j=\gamma_j^2$ ranks a channel last when its gain is small and the channels the model suppresses with a small gain include the massive-activation channels whose true $\mathbb{E}[x_j^2]=\gamma_j^2\,\mathbb{E}[n_j^2]$ is large.
The floor of \S\ref{sec:importance} neutralizes exactly this region: the optimum $f\in[0.055,0.08]$ of Figure~\ref{fig:ablation}b is where the inversion is worst and with the Fisher as importance the floor changes nothing ($0.0315$ at $f=0$ and at $f=0.06$). The per-type rows also say why the oracle's advantage is safe demotion rather than better protection: its bottom $30\%$ holds $1.7\%$ of the mass, \seed{}'s $19.4\%$.

\subsection{Objective versus measured loss}
\label{app:surrogate}

\begin{table}[h]
\centering
\small
\caption{The value of objective~\eqref{eq:objective} at the chosen allocation
(with $e(g)$ in nats, the objective's value is a predicted excess loss) against
the measured excess loss, no outlier columns, at $4.0$ bits per weight unless a
rate is given. ``moved'' is the share of blocks not at the uniform rung.
Llama-2-7B is also shown on the reduced grid of the last three models, as a
dry run (not built).}
\label{tab:surrogate}
\begin{tabular}{llccc}
\toprule
model & grid & moved & objective & measured \\
\midrule
Llama-2-7B   & full ($k\le6$)         & $4.2\%$ & \nop  & 0.0412 \\
Llama-2-7B   & reduced ($k\le4$), dry run & $3.6\%$ & 0.0663 & \nop \\
Llama-2-13B  & reduced                & $3.6\%$  & 0.0579 & 0.0543 \\
Llama-3-8B   & $k\le5$                & $12.4\%$ & 0.1526 & 0.1317 \\
Llama-3-8B, $3.50$ & $k\le5$           & $13.5\%$ & 0.3800 & 0.2886 \\
Llama-3-8B, $3.75$ & $k\le5$           & $23.1\%$ & 0.2368 & 0.1847 \\
Llama-3-8B, $4.25$ & $k\le5$           & $27.2\%$ & 0.1129 & 0.0990 \\
\bottomrule
\end{tabular}
\end{table}

The objective is exact for its inputs and its inputs are approximate: $e(g)$ is
a whole-model average of a rung's error and the sum assumes the blocks'
contributions add. On the five allocations of Llama-2-13B and Llama-3-8B in
the table, which move $4$-$27\%$ of the blocks, the measured loss lands below
the prediction, by $6$-$24\%$.
When most blocks move, as on Qwen2.5-7B, whose hull leaves a narrow window
around $(16,3)$ and which is not among the models of Tables~1--2, the
interaction between rungs that the sum ignores turns the prediction
optimistic.
We therefore never quote the objective as a result.

\subsection{Rung grids}
\label{app:grids}

\begin{table}[h]
\centering
\small
\caption{Uniform rungs built per model and their excess loss $e(g)$ over FP16
(nats per token, WikiText-2 test, no outlier columns). The hull of each column
is the input of \S\ref{sec:alloc}. Rate in bits per weight. ``\nop'': not
built. $^{\ast}$measured but excluded from the allocation.
$^{\dagger}$Build damaged by a storage fault; the arms behind Tables~1--2 are
intact and the fault can only make the Llama-2-13B Mode~1 allocation
suboptimal.}
\label{tab:grids}
\begin{tabular}{lcccccc}
\toprule
rung $(S,k)$ & rate & Llama-2-7B & Llama-2-13B & Llama-3-8B & Mistral-7B & Qwen2.5-7B \\
\midrule
$(8,3)$   & 3.000 & 1.8746 & \nop    & 2.2518$^{\ast}$ & \nop & \nop \\
$(10,3)$  & 3.250 & 0.5852 & $^{\dagger}$ & 0.7014 & 2.5967 & 0.2975 \\
$(12,3)$  & 3.500 & 0.2249 & $^{\dagger}$ & 0.3892 & 0.9774 & 0.1754 \\
$(14,3)$  & 3.750 & 0.1182 & 0.0874 & 0.2445 & 0.3544 & 0.1090 \\
$(16,3)$  & 4.000 & 0.0669 & 0.0583 & 0.1556 & 0.1854 & 0.0744 \\
$(10,4)$  & 3.750 & 0.1445 & \nop    & \nop    & \nop & \nop \\
$(12,4)$  & 4.000 & 0.0744 & \nop    & \nop    & \nop & \nop \\
$(14,4)$  & 4.250 & 0.0441 & 0.0448 & 0.1173 & 0.2457 & 0.0533 \\
$(16,4)$  & 4.500 & 0.0340 & \nop    & \nop    & \nop & \nop \\
$(12,5)$  & 4.500 & 0.0367 & \nop    & \nop    & \nop & \nop \\
$(14,5)$  & 4.750 & 0.0239 & \nop    & \nop    & \nop & \nop \\
$(16,5)$  & 5.000 & 0.0131 & \nop    & 0.0446 & \nop & \nop \\
$(16,6)$  & 5.500 & 0.0064 & \nop    & \nop    & \nop & \nop \\
\bottomrule
\end{tabular}
\end{table}

The grid of Llama-2-7B was built first and in full; the later models were
built with the rungs the time allowed, in the order $(16,3)$, $(14,4)$,
$(14,3)$, $(12,3)$, $(10,3)$, plus $(16,5)$ on Llama-3-8B. The reduced grids
have one rung above $4.0$ bits per weight, $(14,4)$. 
On the full grid of Llama-2-7B the curve the allocator works on is steep and
convex: every two seed bits removed from $S=16$ multiply the excess loss by
$1.8$ to $3.2$ and every coefficient added beyond $k=3$ cuts it by about half
or more, which is what makes a few promotions, paid for by a few demotions,
worth their bits.

\subsection{Per-block error across rungs}
\label{app:rungshape}

Objective~\eqref{eq:objective} assumes that a block's error at rung $g$
factorises as a block term times a rung term, so that one $e(g)$ per rung
serves every block. Table~\ref{tab:rungshape} tests this in weight space on
Llama-3-8B: for $2000$ blocks per tensor in five layers ($50{,}000$ blocks),
the reconstruction error at each rung is divided by the same block's error at
$(16,3)$. The ordering of the rungs holds for the typical block, whose error
ratios are $3.43$, $2.21$, $1.49$, $0.79$ and $0.29$, but single blocks
scatter around it with a log standard deviation of about $0.5$ at every rung,
so that the central $80\%$ of blocks spans a factor of $3.5$-$3.9$. The
scatter has the same size at every rung, consistent with independent variation
in how well the best seed fits a block at each rung. The objective prices
every block at its rung's average and this scatter is one reason the measured
loss departs from it (Appendix~\ref{app:surrogate}).

\begin{table}[h]
\centering
\small
\caption{Ratio of a block's reconstruction error at rung $g$ to its error at
$(16,3)$, Llama-3-8B, $2000$ blocks per tensor in layers $0,7,15,23,31$ of
$\mathbf{W}_q,\mathbf{W}_k,\mathbf{W}_v,\mathbf{W}_{\mathrm{gate}},\mathbf{W}_{\mathrm{up}}$.}
\label{tab:rungshape}
\begin{tabular}{lcccc}
\toprule
rung & median & $q_{10}$ & $q_{90}$ & sd of $\log$ ratio \\
\midrule
$(10,3)$ & 3.43 & 1.75 & 6.62 & 0.54 \\
$(12,3)$ & 2.21 & 1.18 & 4.21 & 0.52 \\
$(14,3)$ & 1.49 & 0.80 & 2.77 & 0.51 \\
$(14,4)$ & 0.79 & 0.41 & 1.49 & 0.52 \\
$(16,5)$ & 0.29 & 0.15 & 0.59 & 0.55 \\
\bottomrule
\end{tabular}
\end{table}

\subsection{S-Quant's rate and our replication}
\label{app:squant}

S-Quant~\citep{wang2026squant} uses SeedLM's generator with an adaptive
coefficient count $k_b$ per block and int8 coefficients under one FP16 scale
per group of $G$ blocks; its rate, $(S+8k_b+16/G)/B$ averaged over blocks,
counts the seed, the coefficients and the scale but not the per-block $k_b$
that a decoder needs to parse the stream. Stored explicitly, $k_b\in\{2,\dots,6\}$
costs $3$ bits per block, $0.19$ bits per weight at the paper's $B=16$, which
puts its $3.8$-bit configuration at about $3.99$ effective bits. Its
comparison rows for SeedLM and the other baselines are transcribed from
\citet{shafipour2025seedlm}. No code is public; we implemented the method from
the paper (the reading reproduces its headline rates: $B=16$, $S=16$, $G=8$,
$R_{\mathrm{th}}=0.90$ gives $3.73$ bits, $R_{\mathrm{th}}=0.80$ gives
$2.63$) and measured weight-space reconstruction error on Qwen2.5-7B and
Llama-3-70B: at $3.8$ bits S-Quant's median relative error is $2.1$-$2.3\times$
SeedLM's at $4.0$ bits and it reaches SeedLM's 4-bit error at $5.2$ bits on
both models. Its adaptive $k_b$ does help within its own format ($8.7\%$
lower error than a fixed $k$ at $3\%$ more bits, on every tensor tested); the
int8 coefficient costs twice SeedLM's 4-bit mantissa and buys $35\%$ fewer
basis vectors at any rate, which is where the difference lives. The perplexities of
S-Quant in Table~\ref{tab:ppl} are the published ones, which correspond to
about $3.99$ effective bits per weight.

\section{Hardware Implementation}
\label{app:seedq_hw}

\begin{figure}[!htbp]
\centering
\includegraphics[width=0.99\linewidth]{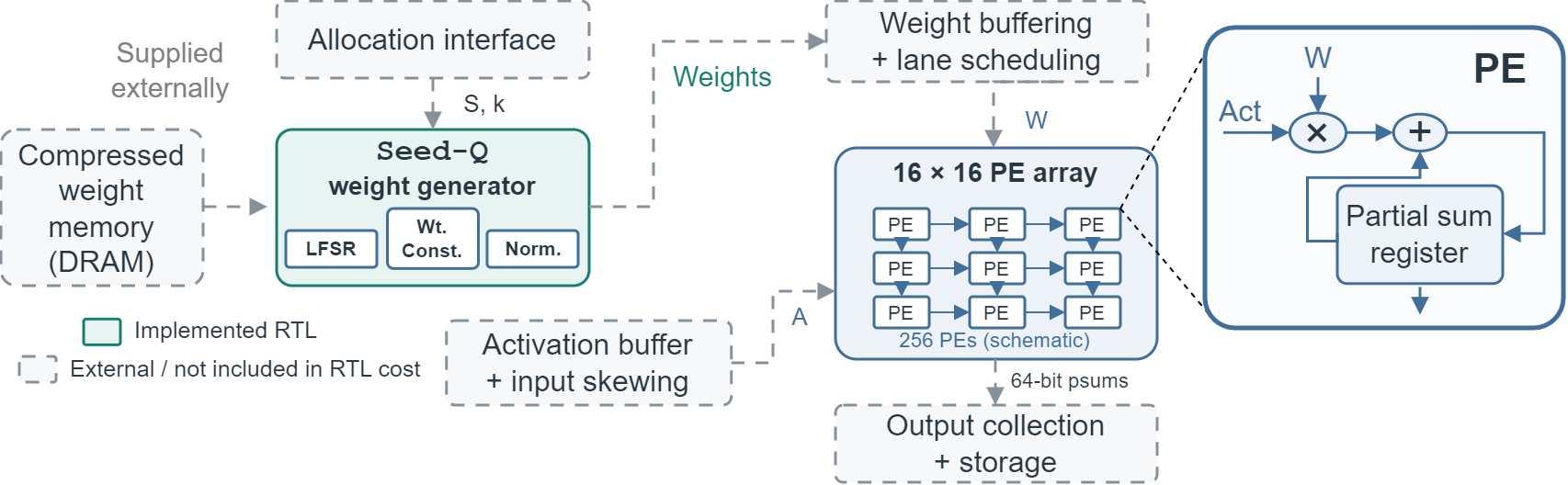}
\caption{\seed~hardware implementation including the weight generator and a shared (16$\times$16) PE array. The generator and the PE array are evaluated separately. Seed (S) and basis count (k) are supplied externally. The common PE multiplies reconstructed weights by activations and accumulates the products in a partial-sum register.}
\label{fig:hardware}
\end{figure}

Figure~\ref{fig:hardware} illustrates the realized hardware implementation of \seed~including the weight generator and its intended integration with a shared PE array. We independently implement SeedLM and S-Quant reconstruction kernels based on their published formulations, using method-specific specializations of a common \texttt{weight\_core} and the same PE design for a fair comparison, while noting that our S-Quant implementation uses direct LFSR generation rather than the published LUT architecture.
\seed accepts a seed and externally supplied allocation parameters: seed width (S$\in [8,16]$) and basis count (k$\in [2,6]$). A configurable LFSR generates states that are centered and multiplied by the coefficients using one reused integer MAC lane. The kernel reconstructs one weight at a time, applies exponent scaling and normalizes the accumulated value, producing signed outputs.

In contrast, SeedLM kernel uses the same coefficient and exponent formats but fixes (S=$16$) and (k=$3$). \seed~adds runtime seed-width and basis-count support, including width-dependent LFSR feedback taps. The S-Quant variant fixes (S=$16$) and supports (k=$\in[2,6]$). All three implementations share the output convention and serial normalization architecture; at (S=$16$) and (k=$3$) configuration, their request interval is $561$ cycles per eight-weight block, as estimated using a 10 ns clock period (100 MHz).

The common PE multiplies activations by reconstructed weights and accumulates sign-extended products in a 64-bit register. A (16$\times$16) array of these PEs is implemented and evaluated separately. Sensitivity estimation, allocation-schedule generation, bitstream unpacking and full memory-to-array scheduling are outside the evaluated reconstruction kernels. The comparison, therefore, characterizes our reconstruction implementations under common arithmetic conventions rather than bit-exact reproductions of the authors’ hardware or a complete integrated accelerator.

\section{Security Evaluation Setup}
\label{app:sec_setup}

This experiment examines one selected fault scenario for each representation
in one tensor. It does not test a physical attack or a deployed integrity
check. We replace the layer-0 query projection $\mathbf W_q$ of Llama-3-8B
($R=C=4096$, 16.78\,M weights) with each scheme's clean or corrupted
reconstruction. All other parameters retain their checkpoint values.
No-Q (BF16) is the same non-quantized 16-bit model denoted \emph{Baseline}
in Tables~\ref{tab:ppl} and~\ref{tab:accuracy}.
WikiText-2 perplexity (PPL)
measures next-token prediction quality, with lower values indicating better
predictions. We evaluate four non-overlapping 2,048-token windows from a
local raw-text file. ARC-Easy accuracy uses 100 sampled test questions,
scored zero-shot by length-normalized answer log-likelihood. Inference runs
on the CPU in BF16 with random seed zero. These measurements isolate one
tensor's fault response from full-model quantization effects.

\paragraph{No-Q (BF16) and AWQ.}
For No-Q, we flip mantissa bit 2 of a sampled BF16 weight. AWQ uses symmetric INT4,
99th-percentile clipping and an FP16 scale per 128 weights, at
4.13 bits/weight. We flip exponent bit 12 of one scale. This implementation
uses AWQ's grouped storage structure but omits activation-aware
calibration~\citep{lin2024awq}.

\paragraph{SeedLM, S-Quant and \seed.}
All three use $B=8$. SeedLM uses rung $g=(16,3)$ and $m=e=4$, at
4 bits/weight, with a simplified LFSR and 256 candidate seeds per block.
We select a seed bit that changes every weight in its block.
S-Quant uses two bases with two coefficients each and SeedLM field widths,
at 4.50 bits/weight. This differs from its published int8 coefficients
and grouped FP16 scales. Since the implementation has no coefficient-count
field, we impose a mismatch halfway through the tensor,
$b_0=1{,}048{,}576$. This is a modeled count error, not a bit flip
in an implemented S-Quant bitstream.

\seed{} uses $S_b=16$, $m=e=4$, $k_b\in\{2,\ldots,6\}$,
and target rate $\bar r=4$ bits/weight. A gain-derived schedule with
slope 1 varies $k_b$ using the RMSNorm gain
$\boldsymbol{\gamma}$ from \eqref{eq:gamma}. This restricted schedule
does not implement the full importance estimate or rung-hull allocation
in \eqref{eq:imp} and \eqref{eq:ladder}; it also omits the full decoder's
cross-tensor dependencies. In this restricted test, a sampled bit
perturbation of $\boldsymbol{\gamma}$ first changes the recomputed
allocation at $b_0=296$. The sampler operates on an FP32 copy of
$\boldsymbol{\gamma}$, although the checkpoint stores BF16. The
successful channel and bit were not logged, so we cannot verify that
this perturbation corresponds to any physically stored BF16 bit. The
full method selects rungs from stored gains and side tables before
decoding; changing a seed does not change later allocations.

For S-Quant and \seed{}, we approximate the consequences of
incorrect block boundaries by substituting decoded blocks with other
stored blocks from the affected tail. We then run model forward passes
to measure PPL and accuracy. These are measured responses to substituted
weights, but the substitutions are not the output of a corrupted packed
decoder. They do not establish that a single stored-bit flip would
produce the reported footprint or model degradation.

\paragraph{Scope.}
This experiment covers one tensor and one selected fault per scheme.
Different bit choices can change the comparison; a BF16 exponent fault
can be destructive even with a single affected weight. Pre-cast footprint
counts can include differences rounded away during BF16 inference.
The S-Quant mismatch begins at the tensor midpoint, whereas the
\seed{} mismatch begins near its start; their different
footprints cannot establish an intrinsic security ranking. Stored-bit
injection, packed-decoder experiments, matched trigger positions and
repeated faults across layers are needed to validate the simulated
amplification. Channel reordering is also unevaluated. The large
degradation suggests a possible monitoring signal, while demonstrating
a risk of service failure in the absence of detection and recovery.

\end{document}

%% file: math_commands.tex
\usepackage{amsmath,amsfonts,bm}

\def\eqref#1{equation~\ref{#1}}

\def\1{\bm{1}}

\DeclareMathAlphabet{\mathsfit}{\encodingdefault}{\sfdefault}{m}{sl}
\SetMathAlphabet{\mathsfit}{bold}{\encodingdefault}{\sfdefault}{bx}{n}

